\documentclass{article}

\usepackage[colorlinks]{hyperref}
\hypersetup{
    colorlinks,
    linkcolor=black,
    citecolor=black,
    filecolor=magenta,
    urlcolor=cyan,
}

\usepackage{resizegather}

\usepackage{amsmath,amssymb,amsfonts,mathrsfs}
\usepackage{mdframed}
\usepackage{graphicx}
\usepackage{textcomp}
\usepackage{xcolor}
\usepackage{tcolorbox}
\usepackage{textcomp}
\usepackage{tablefootnote}
\usepackage{threeparttable}
\usepackage{caption, subcaption}
\usepackage{bbm}
\usepackage{dsfont}
\usepackage{tikz}
\usepackage{graphicx}
\usepackage{tkz-euclide}
\usepackage{algorithm}
\usepackage{algpseudocode}
\usepackage{enumerate}
\usepackage{mathtools}
\usepackage{balance}
\usepackage{graphicx}
\usepackage{amsthm}
\usetikzlibrary{positioning}
\usetikzlibrary{shapes}
\usetikzlibrary{shapes.misc}
\usetikzlibrary{shapes.geometric}
\usetikzlibrary{plotmarks}
\usetikzlibrary{intersections}
\usetikzlibrary{calc}
\usetikzlibrary{fit}
\usetikzlibrary{patterns,tikzmark}
\usetikzlibrary{matrix,decorations.pathreplacing,calc}

\tikzset{cross/.style={cross out, draw, 
         minimum size=2*(#1-\pgflinewidth), 
         inner sep=0pt, outer sep=0pt}}

\newcommand{\state}[0]{x}

\newcommand{\norm}[1]{\left\lVert#1\right\rVert}

\newcommand{\vertiii}[1]{{\left\vert\kern-0.25ex\left\vert\kern-0.25ex\left\vert #1 
    \right\vert\kern-0.25ex\right\vert\kern-0.25ex\right\vert}}

\newmdtheoremenv[linecolor=black, backgroundcolor=lightgray!15, innertopmargin=5pt, innerbottommargin=5pt, skipabove=10pt, skipbelow=10pt]{theorem}{\textbf{Theorem}}

\newmdtheoremenv[linecolor=white, backgroundcolor=lightgray!15, innertopmargin=5pt, innerbottommargin=5pt, skipabove=10pt, skipbelow=10pt]{corollary}{\textbf{Corollary}}[theorem]
\newmdtheoremenv[linecolor=white, backgroundcolor=lightgray!15, innertopmargin=5pt, innerbottommargin=5pt, skipabove=10pt, skipbelow=10pt]{lemma}{\textbf{Lemma}}

\newmdtheoremenv[linecolor=white, backgroundcolor=lightgray!15, innertopmargin=5pt, innerbottommargin=5pt, skipabove=10pt, skipbelow=10pt]{problem}{\textbf{Problem}}

\newtheorem{remark}{Remark}

\newtheorem{definition}{\textbf{Definition}}

\newmdtheoremenv[linecolor=white, backgroundcolor=lightgray!15, innertopmargin=5pt, innerbottommargin=5pt, skipabove=10pt, skipbelow=10pt]{objective}{\textbf{Objective}}

\newmdenv[
    linecolor=white, backgroundcolor=lightgray!15, innertopmargin=5pt, innerbottommargin=5pt, skipabove=10pt, skipbelow=10pt
]{graybox}

\makeatletter
\renewcommand{\fps@figure}{htp}
\renewcommand{\fps@table}{htp}
\makeatother

\usepackage[preprint]{neurips_2025}

\usepackage[utf8]{inputenc} 
\usepackage[T1]{fontenc}    
\usepackage{hyperref}       
\usepackage{url}            
\usepackage{booktabs}       
\usepackage{amsfonts}       
\usepackage{nicefrac}       
\usepackage{microtype}      
\usepackage{xcolor}         

\title{CP-NCBF: A Conformal Prediction-based Approach to Synthesize Verified Neural Control Barrier Functions}

\author{%
  Manan Tayal\thanks{Denotes equal contribution} \\
  Center for Cyber Physical Systems\\
  Indian Institute of Science\\
  Bangalore \\
  \texttt{manantayal@iisc.ac.in} \\
  \And
  Aditya Singh$^*$ \\
  Center for Cyber Physical Systems\\
  Indian Institute of Science\\
  Bangalore \\
  \texttt{adityasingh@iisc.ac.in} \\
  \AND
  Pushpak Jagtap \\
  Center for Cyber Physical Systems\\
  Indian Institute of Science\\
  Bangalore \\
  \texttt{pushpak@iisc.ac.in} \\
  \And
  Shishir Kolathaya \\
  Center for Cyber Physical Systems\\
  Indian Institute of Science\\
  Bangalore \\
  \texttt{shishirk@iisc.ac.in} \\
}

\begin{document}

\maketitle

\begin{abstract}
  Control Barrier Functions (CBFs) offer a powerful framework for enforcing safety in autonomous systems, yet constructing them for general nonlinear dynamics remains a significant challenge. Recent approaches have turned to learning-based techniques, such as neural CBFs (NCBFs), to automate this process. However, verifying the correctness of NCBFs is nontrivial due to inherent approximation errors in learning. In this work, we introduce CP-NCBF, a new framework that applies split-conformal prediction to provide probabilistic safety guarantees for neural CBFs, enabling a tunable balance between robustness to outlier-induced errors in the learned NCBF and the strength of the resulting safety assurances. Unlike prior methods that rely on global Lipschitz constraints—which often hinder scalability and result in overly conservative safety sets—our approach is data-efficient, scalable, and yields less restrictive safety regions. We demonstrate the effectiveness of CP-NCBF through a series of case studies, showing that it produces larger and less conservative safe sets compared to existing verification techniques, while maintaining scalability to high-dimensional dynamical systems.
\end{abstract}

\section{Introduction}
\label{section: introduction}
Ensuring safety is a fundamental challenge in control systems, particularly in autonomous systems where real-time safety guarantees are critical. Safety violations can lead to catastrophic consequences, making it imperative to design control strategies that prioritize safety while maintaining operational efficiency. 
Several paradigms have been proposed for safety-critical control. Safe Reinforcement Learning  incorporates safety specifications as constraints during policy learning, enabling data-driven enforcement but lacking formal guarantees and potentially yielding unsafe behaviors during exploration~\citep{achiam2017constrained, NEURIPS2022_9a8eb202, NIPS2017_766ebcd5}. Hamilton–Jacobi reachability offers worst‑case, formal safety assurances via backward‑reachable tubes, but requires solving the Hamilton–Jacobi–Bellman PDE, whose complexity scales exponentially with system dimensionality~\citep {8263977}. 

Control Barrier Functions (CBFs) offer a principled framework for enforcing safety in control‐affine systems~\citep{Ames_2017, ames2019control}. By embedding safety constraints into real‐time Quadratic Programs (QPs)—solvable at high frequencies using modern optimization solvers—CBFs enable provably safe control across diverse applications, such as adaptive cruise control~\citep{ames2014control}, aggressive aerial maneuvers~\citep{7525253, tayal2023control}, and legged locomotion~\citep{ames2019control, nguyen2015safety}. The efficacy and formal safety guarantees hinge on the choice of CBF employed. Although sum-of-squares programming has been leveraged to synthesize polynomial CBFs \citep{1470374,TOPCU20082669}, such methods remain practical only for systems with low-dimensional state spaces.

To address this, Neural Control Barrier Functions (NCBFs) have emerged as a scalable alternative, leveraging neural networks’ universal approximation properties to model valid CBFs across a broader range of systems. Several approaches have been proposed for the training of NCBFs, including learning from expert demonstrations \citep{9303785}, SMT-based techniques \citep{zhao2020synthesizing,abate2021fossil}, and mixed-integer programming formulations \citep{zhao2022verifying}. Additionally, loss functions tailored for NCBF training have been explored in \citep{dawson2022safe}. However, learning-based CBF synthesis introduces inherent learning errors that can compromise safety. Some existing methods mitigate this issue by performing exact verification on ReLU-activated NCBFs \citep{zhang2023exact, zhang2025seev}, while other methods use robust optimization techniques that enforce Lipschitz continuity constraints on the learned CBF \citep{anand2023formally, tayal2024learning, tayal2024semi}. While these methods improve reliability, they are highly conservative, require dense state-space sampling, and often result in overly restrictive safe regions that hinder system performance.

To overcome these limitations, we introduce Conformal Neural CBF (CP-NCBF), a novel framework that synthesizes formally verified NCBFs with probabilistic guarantees. Our approach leverages split-conformal prediction \citep{angelopoulos2022gentleintroductionconformalprediction}, a statistical framework that provides high-confidence bounds on model predictions, enabling us to quantify and correct learning errors in a sample-efficient manner. Unlike prior methods that require strict Lipschitz constraints and dense sampling for training, our approach allows for training on a limited set of data points while ensuring probabilistic safety guarantees. By systematically refining the learned CBF, CP-NCBF constructs larger and less conservative safe sets by allowing for a direct tradeoff between resilience to safety violations and the probabilistic strength of safety. 

To summarize, the overall contributions of this paper are as follows:
\begin{itemize}
    \item We present a novel framework for safety certification of neural control barrier functions (CBFs) by integrating split-conformal prediction to derive probabilistic guarantees on safety.
    \item We introduce a novel training paradigm for learning robust neural CBFs, equipped with quantifiable probabilistic safety assurances.
    \item Our method enables principled control over CBF conservatism in accordance with task-specific safety requirements, while exhibiting scalability to high-dimensional systems beyond the reach of conventional grid-based approaches.
    \item We empirically validate our framework on benchmark control tasks, demonstrating significant reductions in conservatism and superior scalability compared to existing methods.
\end{itemize}

\nocite{jagtap2020formal, tayal2025physics}

\section{Preliminaries} 
\label{section: preliminaries}
In this section, we first present the system model and definition of safety. We then give the relevant background
and notations for Control Barrier Functions (CBFs).

\subsection{System Model and Safety Definition}
We consider a control-affine nonlinear dynamical system defined by the state $\state \in \mathcal{X} \subseteq \mathbb{R}^n$, the control input $u \in \mathcal{U} \subseteq \mathbb{R}^m$, and governed by the following dynamics: 
\begin{align}
    \dot{\state} = f(\state) + g(\state)u,
\end{align}\label{eq: system_dyn}
where $f: \mathbb{R}^n \to \mathbb{R}^n$ and $g: \mathbb{R}^n \to \mathbb{R}^{n \times m}$ are locally Lipschitz continuous functions. We are given a set $\mathcal{C} \subseteq \mathcal{X}$ that represents the \textit{safe states} for the system. Furthermore, the system is controlled by a Lipschitz continuous control policy $\mu: \mathbb{R}^n \to \mathbb{R}^m$. Our focus lies in ensuring the safety of this dynamical system, which is formally defined as follows: 
\begin{definition}[Safety]
    \label{def:positive-invariance}
   A dynamical system is considered safe if the set, $\mathcal{C} \subseteq \mathcal{X} \subseteq \mathbb{R}^n$, is positively invariant under the control policy, $\mu$, i.e, $x(0)\in \mathcal{C}, u(t) = \mu(x(t))$ $\implies$ $x(t) \in \mathcal{C}, ~\forall~t \geq 0$.
\end{definition}


\subsection{Control Barrier Functions}

Control Barrier Functions~\citep{ames2014control, Ames_2017, ames2019control} are widely used to synthesize control policies with positive invariance guarantees, thereby ensuring system safety. The initial step in constructing a Control Barrier Function (CBF) involves defining a continuously differentiable function $h: \mathcal{X} \to \mathbb{R}$, where the \textit{super-level set} of $h$ corresponds to the safe region $\mathcal{C}$. This leads to the following representation:  
\begin{align}
\label{eq:setc1}
    \mathcal{C} & = \{\state \in \mathcal{X} : h(\state) \geq 0\}, \quad
    \mathcal{X} \setminus \mathcal{C} = \{\state \in \mathcal{X} : h(\state) < 0\}.
\end{align}

The interior and boundary of $\mathcal{C}$ are further specified as:  
\begin{align}
    \text{Int}(\mathcal{C}) & = \{\state \in \mathcal{X} : h(\state) > 0\}, \quad
    \partial\mathcal{C} = \{\state \in \mathcal{X} : h(\state) = 0\}.
\end{align}
The function $h$ qualifies as a valid Control Barrier Function if it satisfies the following definition:

\begin{theorem}[\citep{Ames_2017}]{
\label{definition: CBF definition}
Given a control-affine system $\dot x=f(x)+g(x)u$, the set $\mathcal{C}$ defined by \eqref{eq:setc1}, with $\frac{\partial h}{\partial \state}(\state) \neq 0$ for all $\state \in \partial \mathcal{C}$, the function $h$ is called the Control Barrier Function (CBF) defined on the set $\mathcal{X}$, if there exists an extended \textit{class}-$\mathcal{K}$ function $\kappa$ such that for all $\state \in \mathcal{X}$:
\begin{equation}
\label{eq: lie_derivative}
\begin{aligned}
    \underset{ u \in \mathbb{U}}{\text{sup}}\! \left[\underbrace{\mathcal{L}_{f} h(\state) + \mathcal{L}_g h(\state)u} \iffalse+ \frac{\partial h}{\partial t}\fi_{\dot{h}\left(\state, u\right)} \! + \kappa\left(h(\state)\right)\right] \! \geq \! 0,
\end{aligned}
\end{equation}
where $\mathcal{L}_{f} h(\state) = \frac{\partial h}{\partial \state}f(\state)$ and $\mathcal{L}_{g} h(\state)= \frac{\partial h}{\partial \state}g(\state)$ are the Lie derivatives.}
\end{theorem}

As established in \citep{Ames_2017} and \citep{ames2019control}, any Lipschitz continuous control law $\mu(\state)$ that satisfies the condition $\dot{h} + \kappa(h) \geq 0$ guarantees the system's safety when $x(0) \in \mathcal{C}$. Additionally, if the initial state $x(0)$ lies outside $\mathcal{C}$, this condition ensures asymptotic convergence to the safe set $\mathcal{C}$. 



\subsection{Safe Controller Synthesis using CBFs}

Quite often, we have a reference control policy, $\pi_{ref}(x)$, designed to meet the performance requirements of the system. However, such controllers often lack safety guarantees. To ensure the system meets its safety requirements while preserving performance, the reference controller must be minimally adjusted to incorporate safety constraints. This adjustment can be accomplished using the Control Barrier Function-based Quadratic Program (CBF-QP), described as follows:

\begin{equation}
\begin{aligned}
\label{eq: CBF_QP}
\pi_{safe}(x) &= \min_{u \in \mathbb{U} \subseteq \mathbb{R}^m} \norm{u - \pi_{ref}}^2\\
\quad & \textrm{s.t. } \mathcal{L}_f h(x) + \mathcal{L}_g h(x)u + \kappa \left(h(x)\right) \geq 0.
\end{aligned}
\end{equation}

The CBF-QP framework facilitates the synthesis of a provably safe control policy, $\pi_{safe}(x)$, by leveraging a valid CBF, $h$, while staying close to the reference controller to preserve system performance.

\begin{graybox}
\textbf{Challenges}: A key requirement for synthesizing a safe controller using the CBF-QP formulation is the availability of a valid CBF. Designing a valid CBF by hand is impractical for complex, high-dimensional systems. While techniques like sum-of-squares optimization have been employed to construct polynomial CBFs, their applicability has largely been restricted to systems with low-dimensional state spaces.
\end{graybox}


\section{Problem Formulation}
\label{section: Problem Formulation}
\begin{figure}
    \centering
    \includegraphics[width=1.0\linewidth]{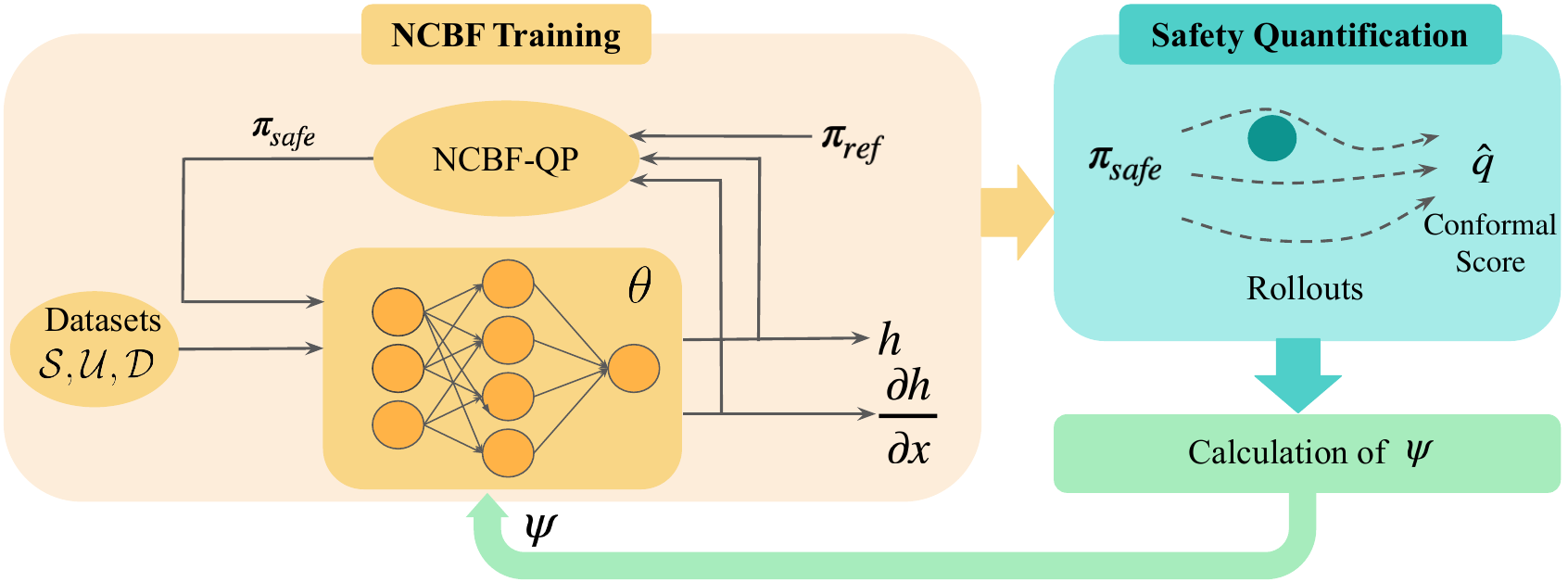}
    \caption{\textbf{Overview of the proposed approach}: The proposed methodology consists of \textbf{three main steps}. In the \textbf{first step}, the NCBF, denoted by $h_{\theta}$, is trained using an initial robustness parameter $\psi = 0$. \textbf{In the second step},  the conformal score $\hat{q}$ is computed based on the policy induced by the learned NCBF. \textbf{In the third step}, this conformal score is then used to update the robustness margin $\psi$, which is subsequently employed to retrain the NCBF. This iterative process is repeated until the conformal score converges to zero, thereby ensuring the desired probabilistic safety guarantee.}
    \label{fig:algo}
\end{figure}

In this section, we formally define the problem of Control Barrier Function (CBF) synthesis and outline the challenges inherent to its direct formulation, particularly those stemming from the need to verify correctness over the entire continuous state space. To overcome these challenges, we propose a reformulation that admits a probabilistic interpretation, enabling formal guarantees on the validity of the solution across the entire state space despite reliance on finitely many data samples.

Leveraging the universal approximation property of neural networks, we introduce a Neural CBF (NCBF) parameterized by a deep neural network (DNN), $h_{\theta}(\state)$, trained via a semi-supervised learning scheme. In the absence of a known CBF, and consequently the true safe set $\mathcal{C}$, we begin with approximate sets: an initial safe set $\mathcal{X}_s \subseteq \mathcal{C}$ and an initial unsafe set $\mathcal{X}_u \subseteq \mathcal{X} \setminus \mathcal{C}$, such that any trajectory originating in $\mathcal{X}_s$ does not enter $\mathcal{X}_u$. Based on this setup, we proceed to formally state the learning problem.

\begin{objective}
\label{prob:origin}
     Given a continuous-time control system defined in Equation~\eqref{eq: system_dyn}, state space $\mathcal{X}$, initial safe and unsafe sets $\mathcal{X}_{s}$ and $\mathcal{X}_{u}$, respectively, the objective is to devise an algorithm to synthesize \text{NCBF} $h_{\theta}(x)$ using a Lipschitz continuous controller $u$ such that 
    \begin{align}
        h_{\theta}(x) \geq 0, \forall x\in \mathcal{X}_s, \quad
        h_{\theta}(x) &< 0,  \forall x\in \mathcal{X}_u, \notag\\
        \frac{\partial h_{\theta}}{\partial x} (f(x)+g(x) u(x)) +  \kappa\left(h_{\theta}\right) &\geq 0,  \forall x\in \mathcal{X}.\label{eq: problem1}
    \end{align}
\end{objective}

In order to enforce conditions \eqref{eq: problem1} in Problem \ref{prob:origin}, we first cast our problem as the following robust optimization problem (ROP):
\begin{equation}\label{eq: rcp}
    \mathrm{ROP}: \begin{cases}\underset{\psi}{\min } & \psi \\ \text { s.t. } & \max \left(q_{k}(x)\right) \leq \psi, k \in\{1,2,3\} \\ & \forall x \in \mathcal{X}, \psi \in \mathbb{R},\end{cases}
\end{equation}

where
\begin{equation} \label{eq:q_conditions}
    \begin{aligned}
        q_{1}(x)=& \left(-h_{\theta}(x)\right) \mathds{1}_{\mathcal{X}_{s}}, \quad
         q_{2}(x)= \left(h_{\theta}(x)+\delta \right) \mathds{1}_{\mathcal{X}_{u}}, \\
         q_{3}(x)=&  -\frac{\partial h_{\theta}}{\partial x} (f(x)+g(x) u(x)) -  \kappa\left(h_{\theta}(x)\right),
    \end{aligned}
\end{equation}
where $\delta$ is a small positive value to ensure the strict inequality. If the optimal solution of the ROP ($\psi_{\mathrm{ROP}}^{*}$) $\leq 0$, then the conditions \eqref{eq: problem1} are satisfied and the $\mathrm{NCBF}$ is valid. 

However, the proposed ROP in \eqref{eq: rcp} has infinitely many constraints since the state of the system is a continuous set. 
This motivates us to employ data-driven approaches and the scenario optimization program of ROP. 

Instead of solving the ROP in \eqref{eq: rcp}, we employ the following scenario optimization problem (SOP):
\begin{equation}\label{eq: scp}
    \mathrm{SOP}: \begin{cases}\underset{\psi}{\min } & \psi \\ \text { s.t. } & q_{1}\left(x_{i}\right) \leq \psi, \forall x_{i} \in \mathcal{S}, \quad q_{2}\left(x_{i}\right) \leq \psi, \forall x_{i} \in \mathcal{U}, \quad q_{3}\left(x_{i}\right) \leq \psi, \forall x_{i} \in \mathcal{D}, \\ & \psi \in \mathbb{R}, i \in\{1, \ldots, N\},\end{cases}
\end{equation}
where $q_{k}(x), k \in\{1,2,3\}$ are defined as in \eqref{eq:q_conditions}. The data sets $\mathcal{S}, \mathcal{U}$ and $\mathcal{D}$ corresponding to points sampled from the initial safe set $\mathcal{X}_s \subseteq \mathcal{C}$, initial unsafe set $\mathcal{X}_u \subseteq \mathcal{X}-\mathcal{C}$, and state set $\mathcal{X}$, respectively.

Given the finite set of data samples $x_i$ and the linear structure of the SOP with respect to the decision variable $\psi$, the optimization problem is feasible. We denote the optimal solution of the SOP, representing a uniform robustness margin, as $\psi^{*}$. Using this premise, we define the primary objective of the paper:
\begin{objective}
\label{obj: Main_obj}
   Given a continuous-time control-affine nonlinear dynamical system and the datasets $\mathcal{S}$, $\mathcal{U}$, and $\mathcal{D}$, our objective is to develop a framework for synthesizing an NCBF $h_{\theta}$ using a uniform robustness margin $\psi$, such that it satisfies the SOP in \eqref{eq: scp} with a user-specified probabilistic confidence level. As a result, the synthesized NCBF and its associated controller ensure safety across the entire state space with the same level of probabilistic guarantee.
\end{objective}

\section{Conformal Neural CBF}
\label{section: cp_cbf}
In this section, we propose an algorithmic approach to solve the problem formulated in Section \ref{section: Problem Formulation}. The structure of this section is as follows. We first present the method to synthesize an NCBF to realize Objective \ref{obj: Main_obj} and then demonstrate the training process.

\subsection{Synthesis of NCBF}

Following the problem formulation in Section~\ref{section: Problem Formulation}, we now present the construction of loss functions for training the NCBF $h_{\theta}(x)$ such that their minimization facilitates the realization of Objective~\ref{obj: Main_obj}. As discussed previously, the NCBF is a feed-forward neural network with trainable weights $\theta$. 

To ensure conformity with the requirements of the SOP in \eqref{eq: scp}, we define the following loss functions over the training datasets $\mathcal{S}$, $\mathcal{U}$, and $\mathcal{D}$:

\begin{equation}\label{eq: ncbf_loss}
    \begin{aligned}
        & \mathcal{L}_{1}(\theta)=\frac{1}{N} \sum_{x_{i} \in \mathcal{S}} \max \left(0,q_1(x_{i}) - \psi^* \right), \quad
        \mathcal{L}_{2}(\theta)=\frac{1}{N} \sum_{x_{i} \in \mathcal{U}} \max \left(0, q_2(x_{i}) - \psi^* \right), \\
        & \mathcal{L}_{3}(\theta)= \frac{1}{N} \sum_{x_{i} \in \mathcal{D}} \max \left(0, q_3(x_{i}) - \psi^* \right), \quad
        \mathcal{L}_{\theta}(\theta)=\mathcal{L}_{1}+\lambda_{1} \mathcal{L}_{2}+\lambda_{2} \mathcal{L}_{3},
    \end{aligned}
\end{equation}

where $\mathcal{L}_{1}$, $\mathcal{L}_{2}$, and $\mathcal{L}_{3}$ penalize violations of the CBF condition in the safe, unsafe, and Lie derivative constraint regions, respectively. The overall loss $\mathcal{L}_{\theta}$ is a weighted combination of these components, with $\lambda_{1}, \lambda_{2} \in \mathbb{R}_{+}$ denoting the relative importance of each term. This formulation enables the systematic training of the NCBF with the robustness margin, $\psi^*$.

\subsection{Conformal Neural CBF}
From Equation~\eqref{eq: ncbf_loss}, it is evident that the choice of $\psi$ plays a critical role in shaping the learned neural CBF. Specifically, setting $\psi > \psi^*$ can lead to safety violations, thereby rendering the CBF incomplete, whereas choosing $\psi \ll \psi^*$ introduces unnecessary conservatism in the control policy. A fundamental challenge, therefore, lies in determining an appropriate value for $\psi^*$. Intuitively, $\psi^*$ represents the worst-case safety violation across the entire state space, implying that accurate estimation of $\psi^*$ requires computing a tight upper bound on the maximum violation over an infinite
verification set. However, evaluating the system over infinitely many state-space points is computationally intractable.

To overcome this limitation, we propose a split conformal prediction-based framework that enables probabilistic upper bounds on the maximum safety violation using only a finite verification set. The following theorem formalizes this finite-sample guarantee:
\vspace{5em}

\begin{theorem}[Safety quantification of Neural CBF] \label{thm:conformal_ncbf} \textit{Consider a continuous time control system \eqref{eq: system_dyn} for which we obtain a Neural CBF $h_{\theta}(\state)$ parameterized by $\theta$. We sample $N$ i.i.d. samples from $\mathcal{X}$. For a user-specified level $\alpha$, let $\hat{q}$ be the $\frac{\lceil(N+1)(1-\alpha)\rceil}{N}th$ quantile of the conformal scores, $s_j = \max_{i \in \{1,2,3\}}(q_i(x_j))$, on the $N$ state samples. Select a safety violation parameter $\epsilon \in (0,1)$ and a confidence parameter $\beta \in (0,1)$ such that:}
\begin{equation} \label{eq: eps_calc}
     \mathcal{I}_{1-\epsilon}(N-l+1, l) \leq \beta,
\end{equation}
\textit{where $l = \lfloor (N+1)\alpha \rfloor$ and $\mathcal{I}$ is the regularized incomplete beta function.
Then, with the probability of at least $1 - \beta$, the following holds:}
\begin{equation}\label{eq: safety_result}
\underset{x \in \mathcal{X}}{\mathbb{P}}\left(\max_{i\in \{1,2,3\}}(q_i(x)) \leq \hat{q} \right) \geq 1- \epsilon.
\end{equation}
\end{theorem}

The proof is available in Appendix \ref{appendix: safety_quantification}. For a sufficiently high confidence level ($1-\beta$) and safety level ($1-\epsilon$), $\hat{q}$ provides a means to quantify the safety of a neural CBF. A positive $\hat{q}$ indicates the presence of safety violations, with its magnitude quantifying the severity of these violations. The procedure to quantify this safety is detailed in Algorithm \ref{alg:CP-NCBF}.

\begin{algorithm}[t]
\caption{Safety Quantification of Neural CBF}
\label{alg:CP-NCBF}
\begin{algorithmic}
\Require $f, g$ (dynamics), $N$, $\alpha$.
   \State $D \gets \text{Sample $N$ IID states from } \{x : x \in \mathcal{X}\}$
    \For{$i = 0, 1, \dots, N-1$}
        \State $S_i \gets s_i(0, D)$
    \EndFor
\State $S \gets S~\text{sorted in decreasing order}$ 
\State $l \gets \lfloor (N+1)\alpha \rfloor$
\State $\beta, \epsilon \gets N, l$ \Comment{From eq. \eqref{eq: eps_calc}}
\State $\hat{q} \gets S_{l}, \beta, \epsilon$ \Comment{From eq. \eqref{eq: safety_result}}
\State \textbf{return} $\hat{q}, \beta, \epsilon$
\end{algorithmic}
\end{algorithm}

\begin{remark}
While the above algorithm offers a probabilistic upper bound on the magnitude of safety violations, it does not yield a direct estimate of the value of $\psi^*$.
\end{remark}

In the subsequent subsection, we detail the procedure to estimate $\psi^*$ from the conformal score $\hat{q}$ along with the training scheme to synthesize a probabilistically verified NCBF.

\subsection{Training Procedure}

We present the training methodology for synthesizing a Neural Control Barrier Function (NCBF) with probabilistic
safety guarantees. At each time step, the controller receives the current system state $x$ and a nominal control input $u_{\text{ref}}(x)$. The goal is to compute a safe control input $u$ that minimally deviates from $u_{\text{ref}}$ while satisfying the learned NCBF-based safety constraint. This is achieved by solving a Quadratic Program (NCBF-QP) that filters out unsafe actions. The resulting control input $u$ is then utilized in computing the training loss for the subsequent optimization step.

The training dataset is partitioned into mini-batches, and the loss is evaluated batch-wise. Optimization is performed using a stochastic gradient descent method such as ADAM to update the learnable parameter $\theta$. The objective is to learn a parameterization such that the conformal score of the trained NCBF is non-positive, ensuring probabilistic safety.

To this end, we employ an iterative refinement strategy. Initially, the NCBF is trained with $\psi = 0$, as initializing with a nonzero $\psi$ typically induces excessive conservatism. The conformal score $\hat{q}$ of the trained NCBF is then evaluated. If $\hat{q} \leq 0$, the number of violations lies within the user-specified threshold, and $\psi$ can remain zero. Otherwise, $\psi$ is updated as $\psi := -\hat{q}$ to directly counteract the violations. The NCBF is then retrained with this new $\psi$. This process is repeated until the conformal score satisfies $\hat{q} \leq 0$ or a predefined maximum number of training iterations is reached. All training is performed offline within a simulated environment. Upon convergence, the resulting NCBF-QP-based controller $u$ can be deployed for online execution on the target system.

This approach provides a scalable and principled way to trade off conservatism with operational flexibility. 
We detail the procedure to evaluate $\psi$ in Algorithm~\ref{alg:CP-Ver}.
\begin{algorithm}[t]
\caption{Training Robust Neural CBFs with probabilistic safety assurances}
\label{alg:CP-Ver}
\begin{algorithmic}[1]
\Require Data Sets: $\mathcal{S}, \mathcal{U}, \mathcal{D}$, Dynamics: $f, g$, $N$, $\alpha$
\State $x_i \gets \text{Sample}(\mathcal{S}, \mathcal{U},\mathcal{D})$
\State $\psi \gets 0$ 
\State Obtain $\mathcal{L}_{\theta}(\theta)$. \Comment{From eq. \eqref{eq: ncbf_loss}} 
\While{$\mathcal{L}_{\theta}(\theta) > 0$}
    \State $h_{\theta} \gets \theta$
    \State $\mathcal{L}_{\theta}(\theta) \gets (h_{\theta}, f, g, x_i, \psi)$ 
    \State $\theta \gets \text{Learn}(\mathcal{L}_{\theta}, \theta)$
\EndWhile
\State Compute $\hat{q}(N$, $\alpha)$ using Algorithm~\ref{alg:CP-NCBF}.
\State $\psi \gets -\hat{q}$ 
\State Obtain $\mathcal{L}_{\theta}(\theta)$ using $\psi$. \Comment{From eq. \eqref{eq: ncbf_loss}} 
\While{$\mathcal{L}_{\theta}(\theta) > 0$}
    \State $h_{\theta} \gets \theta$
    \State $\mathcal{L}_{\theta}(\theta) \gets (h_{\theta}, f, g, x_i, \psi)$ 
    \State $\theta \gets \text{Learn}(\mathcal{L}_{\theta}, \theta)$
\EndWhile
\State \textbf{return} $h_{\theta}$
\end{algorithmic}
\end{algorithm}

\section{Experiments}
\label{section: experiments}
\begin{figure*}[t]
    \centering
    \includegraphics[width=\linewidth]{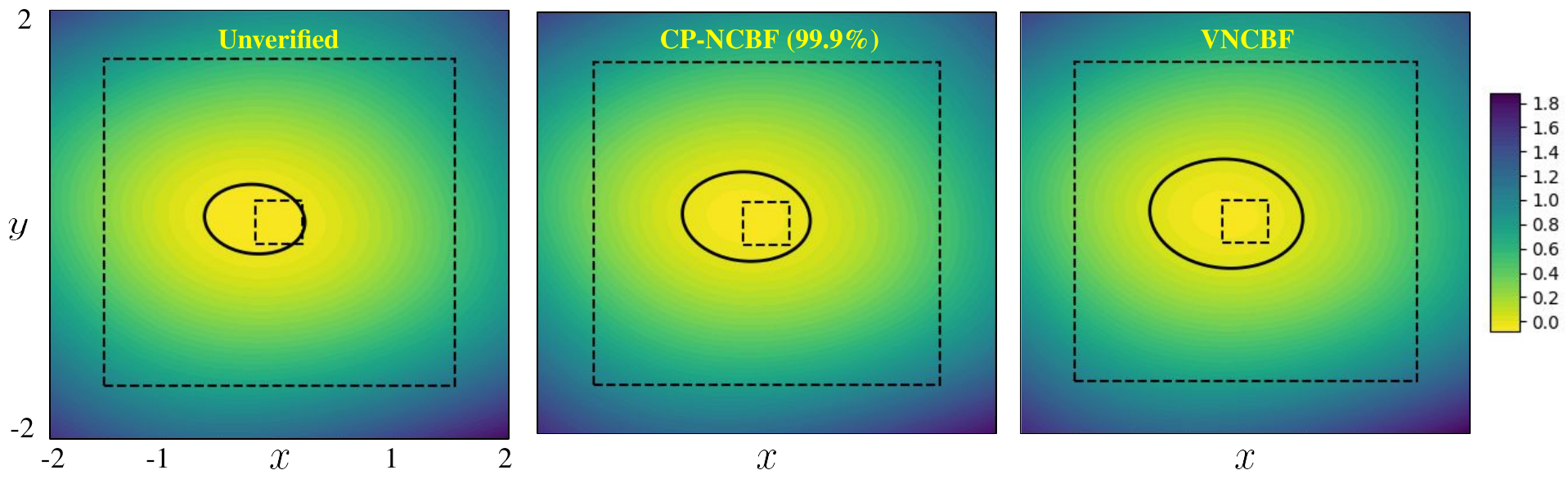}
    \vspace{-0.5em}
    \caption{Visualization of learned NCBFs for autonomous ground vehicle at $\phi=0$: (Left) Unverified NCBF (without robust loss), (Center) CP-NCBF (our method with 99.9\% safety guarantee), and (Right) Verified NCBF \citep{tayal2024learning}. Bold black ovals denote the zero-level sets. Both \textbf{CP-NCBF and Verified NCBF avoid overlapping the unsafe region, while the Unverified NCBF does not}; additionally, CP-NCBF yields \textbf{a larger safe region} than the Verified NCBF}
    \label{fig:uni_comparison_baseline}
\end{figure*}

We evaluate the effectiveness of our proposed framework on four representative systems: (i) autonomous ground vehicle collision avoidance, (ii) aerial vehicle geofencing, and (iii) quadruped navigation in dynamic environments. Our method is benchmarked against the verification strategy introduced in~\citep{tayal2024learning}, which employs Lipschitz regularization and dense sampling to synthesize a verified NCBF. We also compare with an unverified NCBF trained following the approach in~\citep{dawson2022safe}, in order to investigate the impact of verification. Furthermore, we analyze the alignment between the probabilistic safety guarantees provided by our method and the observed empirical safety rates during deployment, thereby validating the reliability of the proposed guarantees.


\subsection{Experimental Case Studies}

\begin{itemize}
    \item \textbf{Autonomous Ground Vehicle Collision Avoidance:} In our first experiment, we examine a $3$-dimensional collision avoidance problem involving an autonomous ground vehicle governed by Dubins' car dynamics~\citep{dubins1957curves}. The objective is to ensure safety by avoiding a static obstacle while navigating through a bounded environment. Further details on the system dynamics, state space bounds, and experimental setup are provided in Appendix~\ref{appendix: Dubins}.
    \item \textbf{Geo-Fencing of Autonomous Aerial Vehicle:} In our next experiment, we investigate a $6$-dimensional aerial geo-fencing scenario, wherein an autonomous aerial vehicle is tasked with following a reference trajectory while remaining within a predefined geo-fence region. The safety constraint enforces that the vehicle's state remains confined within this designated region throughout its operation. Further details on the system dynamics, state space bounds, experimental setup and simulation results are provided in Appendix~\ref{appendix: Geofencing}.
    \item \textbf{Quadruped navigating in a dynamic environment:} In our final experiment, we evaluate an $8$-dimensional quadruped navigation task, wherein a quadruped robot must safely navigate an environment while avoiding a dynamic obstacle. This scenario introduces significant complexity due to the time-varying nature of the obstacle, which poses challenges in enforcing safety constraints and increases the dimensionality of the underlying dynamical system. Moreover, this setting serves as a representative benchmark to assess the scalability of our proposed framework to high-dimensional, safety-critical control tasks. Further details on the system dynamics, state space bounds, experimental setup and simulation results are provided in Appendix~\ref{appendix: quadruped}.
\end{itemize}

\begin{figure*}[t]
    \centering
    \includegraphics[width=0.95\linewidth]{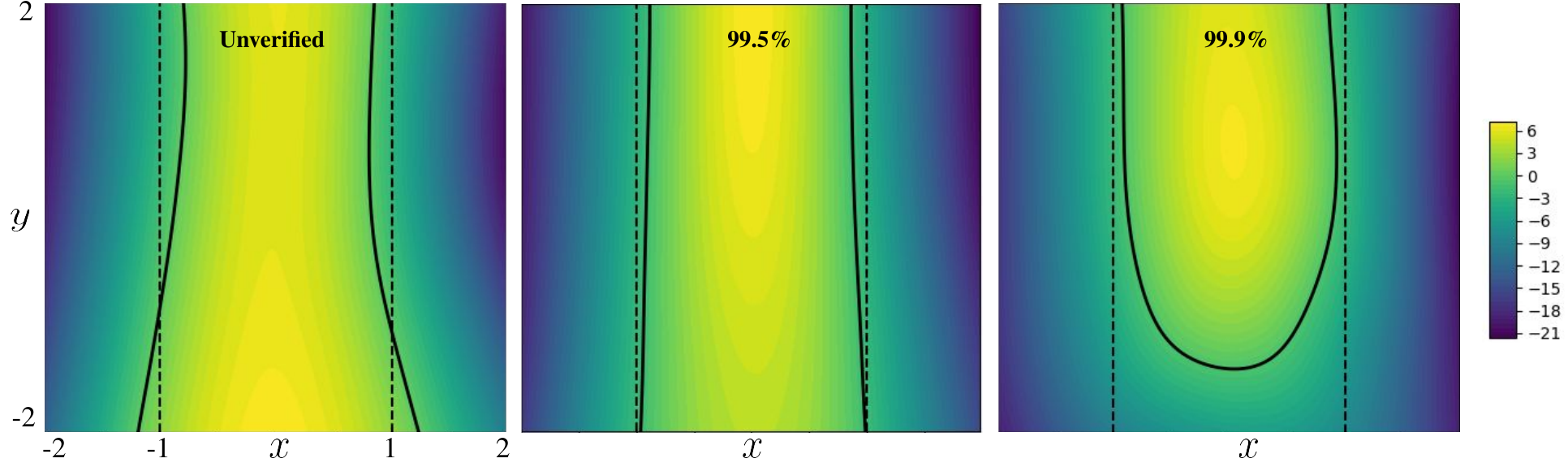}
\vspace{-0.5em}
\caption{Visualization of learned NCBFs for the aerial vehicle at $\phi, \dot{x}_1, \dot{x}_2, \dot{\phi} = 0$: (Left) Unverified NCBF (without robust loss), (Center) CP-NCBF (with \textbf{99.5\% safety guarantee}), and (Right) CP-NCBF (with \textbf{99.9\% safety guarantee}). Bold black lines denote the zero-level sets. As the desired \textbf{safety guarantees become more stringent}, the learned NCBF \textbf{exhibits increased conservatism}. This behavior highlights the flexibility of our proposed framework in enabling a \textbf{systematic trade-off between safety and performance, tailored to the specific requirements of the system.}}
\label{fig: comp_safety_levels}
\vspace{-1em}
\end{figure*}


\begin{figure}
    \centering
    \includegraphics[width=0.9\linewidth]{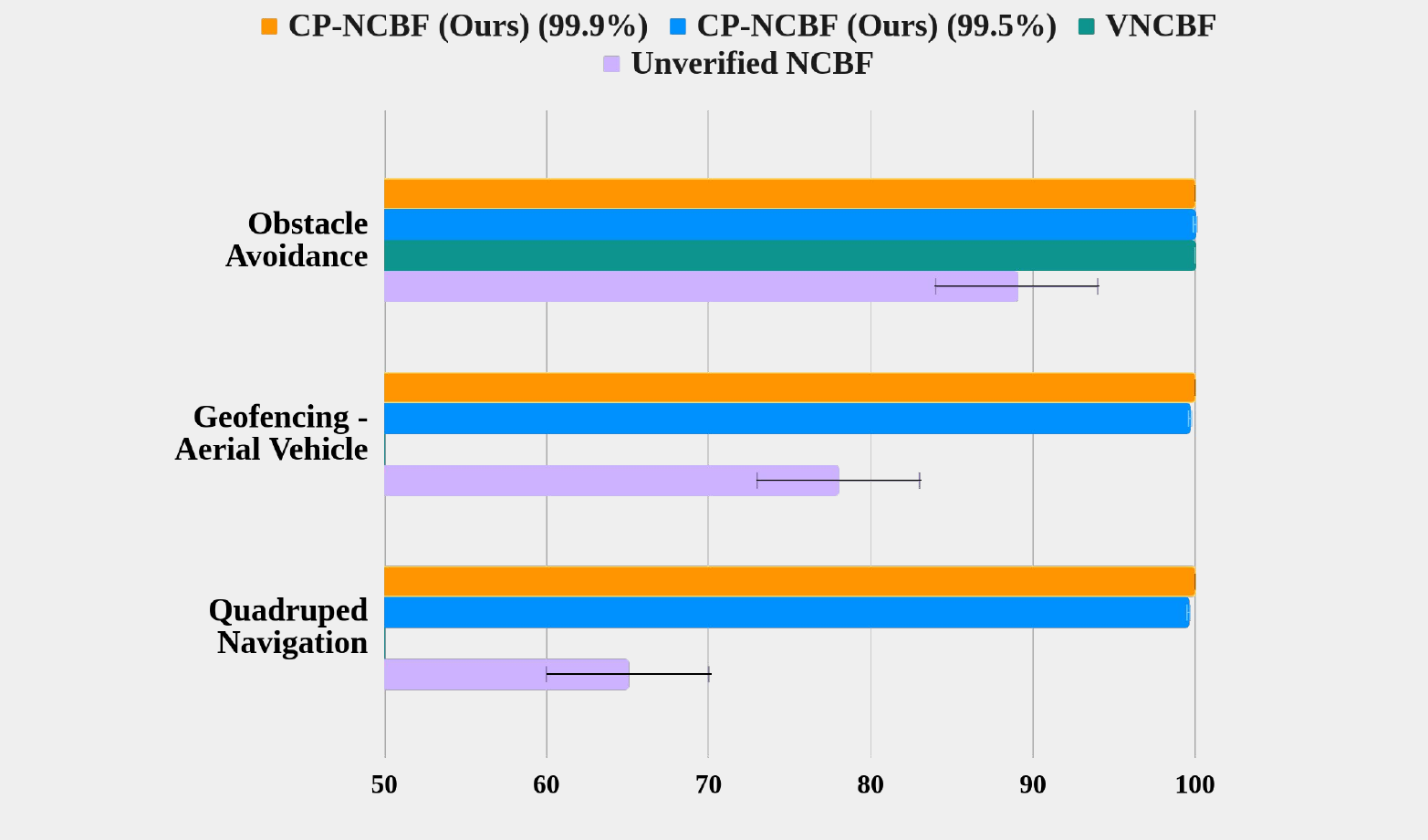}
    \caption{This figure presents a comparative analysis of the empirical safety rates achieved by all evaluated methods. \textbf{The observed safety rates align closely with the corresponding theoretical guarantees}, thereby providing empirical validation of the reliability and robustness of the proposed framework. These findings underscore the potential of our approach for \textbf{deployment in real-world safety-critical applications}. Furthermore, we note that \textbf{VNCBF could not be computed} for the geo-fencing and quadruped navigation tasks due to its \textbf{reliance on grid-based sampling, which is computationally feasible only for systems up to $5$ dimensions}. This limitation further illustrates the \textbf{scalability advantage of our method, which remains tractable and effective even in high-dimensional settings.}} 
    \label{fig: emp_safety}
\end{figure}

\subsection{Comparative Analysis} 

Figure~\ref{fig:uni_comparison_baseline} presents a comparative analysis of the zero-level sets for the autonomous ground vehicle collision avoidance task at $\phi = 0$, across three approaches: (left) an unverified
NCBF trained without robustness considerations, (center) the proposed Conformal NCBF (CP-NCBF) offering a $99.9\%$ probabilistic safety guarantee, and (right) a verified NCBF synthesized using the method of~\citep{tayal2024learning}. The unverified NCBF exhibits overlap with the predefined unsafe set, revealing safety violations likely induced by insufficient generalization. In contrast, both the CP-NCBF and the verified NCBF avoid unsafe regions, highlighting the necessity of incorporating verification during NCBF synthesis. 

A similar trend is observed in the aerial vehicle geo-fencing task depicted in Figure~\ref{fig: comp_safety_levels}, further validating the efficacy of safety-aware training. Importantly, the CP-NCBF yields a considerably less conservative approximation of the safe set compared to~\citep{tayal2024learning}, while still maintaining high safety levels ($99.9\%$), thereby demonstrating an improved trade-off between conservatism and performance. Moreover, Figure~\ref{fig: comp_safety_levels} illustrates that tightening the safety guarantees leads to increasingly conservative safe regions. This characteristic enables practitioners to systematically adjust the degree of conservatism to suit application-specific safety-performance trade-offs. Thus, our framework facilitates flexible and principled tuning of control policies under safety constraints.

These findings generalize to higher-dimensional tasks as well. Figure~\ref{fig: quad_nav} (Appendix~\ref{appendix: quadruped}) illustrates similar trends in the quadruped navigation setting, validating the scalability and practical relevance of our framework for deployment in complex, high-dimensional, safety-critical systems.

Lastly, Figure~\ref{fig: emp_safety} summarizes the empirical safety rates attained by all evaluated methods, aggregated over five random seeds with error bars representing the corresponding variance. Notably, the VNCBF baseline could not be computed for the geofencing and quadruped navigation tasks due to computational intractability, thereby reinforcing our claim regarding the scalability advantages of the proposed verification framework in high-dimensional settings. Furthermore, the observed empirical safety rates are in close agreement with the theoretical guarantees, highlighting both the reliability and practical applicability of our probabilistic verification framework in safety-critical autonomous systems.


\section{Conclusion}
\label{section: conclusions}
We propose CP-NCBF, a framework for synthesizing neural CBFs with probabilistic safety guarantees via split conformal prediction. Our method bridges the gap between learning-based synthesis and formal safety verification, enabling the construction of less conservative safe sets with improved sample efficiency. Case studies on autonomous driving and aerial vehicle geo-fencing tasks validate that our method achieves real-time safety and broadens the safe operational domain compared to existing approaches.

\textbf{Limitations and Future Work:} The current work assumes the knowledge of the dynamics ($f$ and $g$). Moreover, the methodology currently only focuses on safety, without any focus on performance objectives. As a part of future work, we plan to extend the framework to account for unknown dynamics and also work on synthesizing NCBFs to co-optimize both safety and performance. It would also be exciting to extend this framework to more complex, high-dimensional
systems.

\bibliographystyle{plainnat}
\bibliography{ref}
\newpage
\appendix
\section{Theorem \ref{thm:conformal_ncbf}} \label{appendix: safety_quantification}

\begin{tcolorbox}[colback=gray!10, colframe=gray!80, boxrule=0.5pt, arc=4pt]
\textbf{Theorem \ref{thm:conformal_ncbf}} (Safety quantification of Neural CBF) \textit{Consider a continuous time control system \eqref{eq: system_dyn} for which we obtain a Neural CBF $h_{\theta}(\state)$ parameterized by $\theta$. We sample $N$ i.i.d. samples from $\mathcal{X}$. For a user-specified level $\alpha$, let $\hat{q}$ be the $\frac{\lceil(N+1)(1-\alpha)\rceil}{N}th$ quantile of the conformal scores, $s_j = \max_{i \in \{1,2,3\}}(q_i(x_j))$, on the $N$ state samples. Select a safety violation parameter $\epsilon \in (0,1)$ and a confidence parameter $\beta \in (0,1)$ such that:}
\begin{equation*}
     \mathcal{I}_{1-\epsilon}(N-l+1, l) \leq \beta,
\end{equation*}
\textit{where $l = \lfloor (N+1)\alpha \rfloor$ and $\mathcal{I}$ is the regularized incomplete beta function.
Then, with the probability of at least $1 - \beta$, the following holds:}
\begin{equation*}
\underset{x \in \mathcal{X}}{\mathbb{P}}\left(\max_{i\in \{1,2,3\}}(q_i(x)) \leq \hat{q} \right) \geq 1- \epsilon.
\end{equation*}
\end{tcolorbox}

\begin{proof}
To lay the foundation for our methodology, we first introduce Split Conformal Prediction:
\begin{lemma}[Split Conformal Prediction \citep{angelopoulos2022gentleintroductionconformalprediction}]
\label{lem:split_conformal}
Consider a set of independent and identically distributed (i.i.d.) calibration data, denoted as $\{(X_i, Y_i)\}_{i=1}^n$, along with a new test point $(X_{\text{test}}, Y_{\text{test}})$ sampled independently from the same distribution. Define a score function $s(x, y) \in \mathbb{R}$, where higher scores indicate poorer alignment between $x$ and $y$. Compute the calibration scores $s_1 = s(X_1, Y_1), \ldots, s_n = s(X_n, Y_n)$. For a user-defined confidence level $1-\alpha$, let $\hat{q}$ represent the $\lceil (n+1)(1-\alpha) \rceil / n$ quantile of these scores. Construct the prediction set for the test input $X_{\text{test}}$ as:
\begin{equation*}
\mathcal{C}(X_{\text{test}}) = \{y : s(X_{\text{test}}, y) \leq \hat{q} \}.
\end{equation*}
Assuming exchangeability, the prediction set $\mathcal{C}(X_{\text{test}})$ guarantees the marginal coverage property:
\begin{equation*}
\mathbb{P}(Y_{\text{test}} \in \mathcal{C}(X_{\text{test}})) \geq 1 - \alpha.
\end{equation*}
\end{lemma}

\begin{remark}
\textit{The coverage of conformal prediction, conditioned on the calibration set, is a random variable. Running the algorithm with different calibration sets results in varying coverage over an infinite validation set. While Lemma \ref{lem:split_conformal} ensures coverage is at least $1 - \alpha$ on average, it deviates for any fixed calibration set.}
\textit{However, increasing $n$ reduces these fluctuations, and the coverage distribution follows an analytic form, as derived by Vladimir Vovk in \citep{vovk2012}:}
    \begin{equation}
    \label{eq: beta_coverage}
        \mathbb{P}(Y_{\text{test}} \in \mathcal{C}(X_{\text{test}})) \sim \text{Beta}(n + 1 - l, l), 
    \end{equation}
   \textit{ where $l = \lfloor (n+1)\alpha \rfloor$}.
\end{remark}

    Using Lemma \ref{lem:split_conformal}, we define the input as $ X = x $ and the corresponding output as $Y = \max_{i \in \{1,2,3\}} (q_i(x))$.  
The conformal scoring function is then given by $s(X, Y) := \max_{i \in \{1,2,3\}} (q_i(x))$,
where higher scores correspond to greater misalignment between $ x $ and $ y $, or equivalently, indicate higher safety violations in our setting.  

Accordingly, the prediction set for a test input $ X_{\text{test}} = x $ is defined as:  
    \begin{equation*}
        \mathcal{C}(X_{test} = x) = \{y: \max(q_i(x)) \leq \hat{q}\},~\forall x \in \mathcal{X}.
    \end{equation*}
    Now, using eq \eqref{eq: beta_coverage}, we get
    \begin{equation*}
        \underset{x \in \mathcal{X}}{\mathbb{P}}\left(\max_{i \in \{1,2,3\}}(q_i(x)) \leq \hat{q} \right) \sim \text{Beta}(N +1 - l, l).
    \end{equation*}
    Define $ E $ as:
\begin{equation*}
E := \underset{x \in \mathcal{X}}{\mathbb{P}}\left(\max_{i \in \{1,2,3\}}(q_i(x)) \leq \hat{q} \right).
\end{equation*}
Here, $E$ is a Beta-distributed random variable. Using properties of CDF, we assert that $ E \geq 1 - \epsilon $ with confidence $ 1 - \beta $ if the following condition is satisfied:
\begin{equation} \label{eq:eps_calc_perf}
    \mathcal{I}_{1-\epsilon}(N - l + 1, l) \leq \beta,
\end{equation}
where $\mathcal{I}_x(a,b)$ is the regularized incomplete Beta function and also serves as the CDF of the Beta distribution.

Thus, if Equation~\eqref{eq:eps_calc_perf} holds, we can conclude with probability $ 1-\beta $ that:
\begin{equation*}
\underset{x \in \mathcal{X}}{\mathbb{P}}\left(\max_{i \in \{1,2,3\}}(q_i(x)) \leq \hat{q} \right) \geq 1-\epsilon.
\end{equation*}
\end{proof}

\section{Relationship between \texorpdfstring{$\alpha$,~$\beta$, and $\epsilon$}{alpha, beta, and epsilon}}
\begin{figure}[h]
\centering
\includegraphics[width=1.0\linewidth]{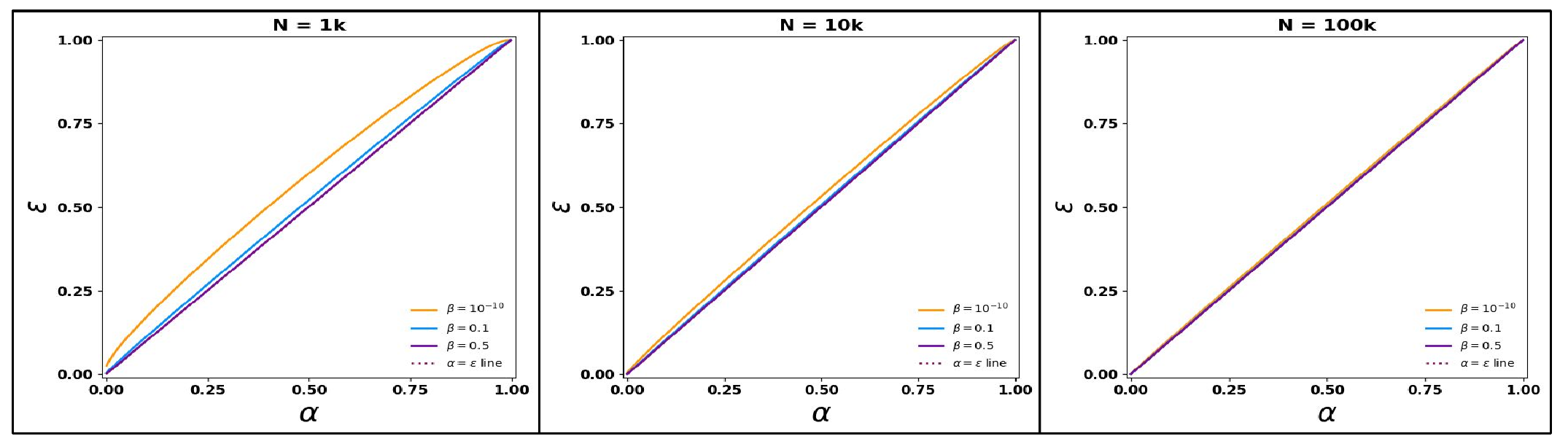}
\caption{This figure shows the $\alpha$-$\epsilon$ plots for different numbers of verification samples, $N$, and different values of $\beta$.}
\label{fig: cp_analysis}
\end{figure}

The work~\citep{vovk2012} states that a smaller number of samples leads to greater fluctuations in the conformal prediction calibration, meaning that if we redraw $N$ samples and repeat the conformal prediction process, we might get a different calibration result. This variance decreases as $N$ increases.Similarly, in our work, a small $N$ means that the value correction term $\delta$ might fluctuate each time the verification algorithm is executed. Therefore, to ensure a stable estimate of $\delta$, it is desirable to select a sufficiently large value of $N$.

Figure~\ref{fig: cp_analysis} presents the $\alpha-\epsilon$ plots for varying numbers of verification samples $N$ and different values of $\beta$. From the figure, we observe that as $N$ increases, the effect of $\beta$ diminishes, and the curve approaches the $\alpha = \epsilon$ line. Ideally, the user-specified safety error rate ($\alpha$) should closely match the safety violation parameter ($\epsilon$) while maintaining high confidence ($1-\beta$ close to 1).  
Thus, selecting a larger $N$ enables a smaller $\beta$ while ensuring the alignment of $\alpha$ and $\epsilon$. Conversely, if $N$ is small, one must either compromise on the confidence parameter $\beta$ or accept that $\alpha$ will be lower than $\epsilon$, resulting in a more conservative upper bound on the safety rate.

\section{Experimental Details}

\subsection{Autonomous Ground Vehicle Collision Avoidance}\label{appendix: Dubins}
We consider an autonomous ground vehicle collision avoidance problem with the state space given by $\mathcal{X}= [-2,2]^2 \times [-\pi, \pi]$. The safe region given by $\mathcal{X}_{s}=\mathcal{X} \backslash [-1.5, 1.5]^{2} \times [-\pi, \pi]$ and the unsafe region by $\mathcal{X}_{u}=[-0.2, 0.2]^{2} \times [-\pi, \pi]$.

\subsection{Aerial Vehicle Geo-Fencing}\label{appendix: Geofencing}

\begin{figure}
    \centering
    \includegraphics[width=0.6\linewidth]{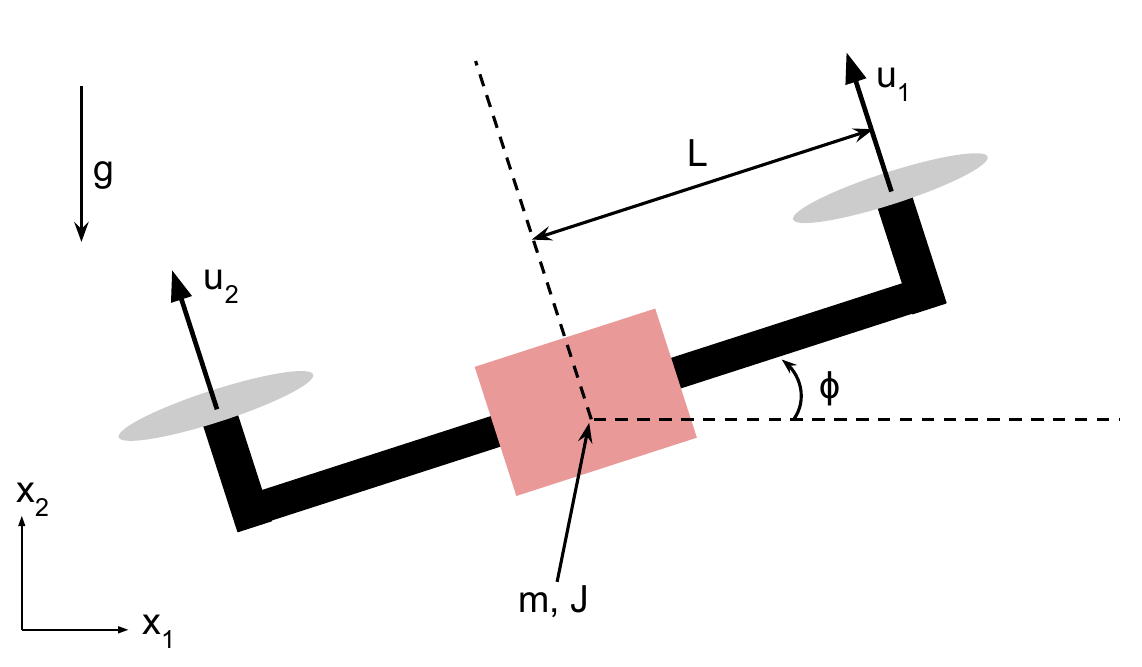}
    \caption{A visual representation of the aerial vehicle}
    \label{fig:planar-uav}
\end{figure}

The control-affine dynamics of a planar aerial vehicle are given as:
\begin{equation}
    \left[\begin{array}{l}
    \ddot{x}_1 \\
    \ddot{x}_2 \\
    \ddot{\phi}
    \end{array}\right]=\left[\begin{array}{c}
    0 \\
    -g \\
    0
    \end{array}\right]+\begin{bmatrix}
        -\frac{1}{m}\sin\phi & -\frac{1}{m}\sin\phi \\
        \frac{1}{m}\cos\phi & \frac{1}{m}\cos\phi \\
        \frac{1}{LJ} & -\frac{1}{LJ}
    \end{bmatrix}
    \left[\begin{array}{c}
    u_1 \\
    u_2
    \end{array}\right],
\end{equation}
where $[x_1, x_2, \phi, \dot{x}_1, \dot{x}_2, \dot{\phi}]^T\in\mathcal{X}\subseteq\mathbb{R}^6$ represents the state vector comprising the position $(x_1, x_2)$ and orientation $\phi$ of the vehicle. The inputs $u_1$ and $u_2$ correspond to the thrusts generated by the right and left motors, respectively, with mass $m$, moment of inertia $J$, and motor separation length $2L$. The gravitational constant is $g$. We normalize the parameters to $m = 1$ and $1/LJ = 1$ for simplicity.

The state space is defined as $\mathcal{X}= [-2,2]^2 \times [-\pi, \pi] \times [-2,2]^3$. The safe region is $\mathcal{X}_{s} = [-0.8, 0.8]^{2} \times [-\pi, \pi] \times [-2,2]^3$, while the unsafe region is $\mathcal{X}_{u} = \mathcal{X} \backslash [-1, 1]^{2} \times [-\pi, \pi] \times [-2,2]^3$. 

\subsubsection{Simulation Results}
We evaluated the CP‑NCBF‑QP controller (Eq.~\eqref{eq: CBF_QP}) at a $99.5\%$ safety level in a UAV environment built on the Pybullet physics engine \citep{coumans2019} using the Pybullet‑Drones framework \citep{pybullet-drones}. The vehicle is required to track a predefined reference trajectory while ensuring it remains within the geofenced interval $x_1 \in [-1,1]$.

\begin{figure}
    \centering
    \includegraphics[width=0.6\linewidth]{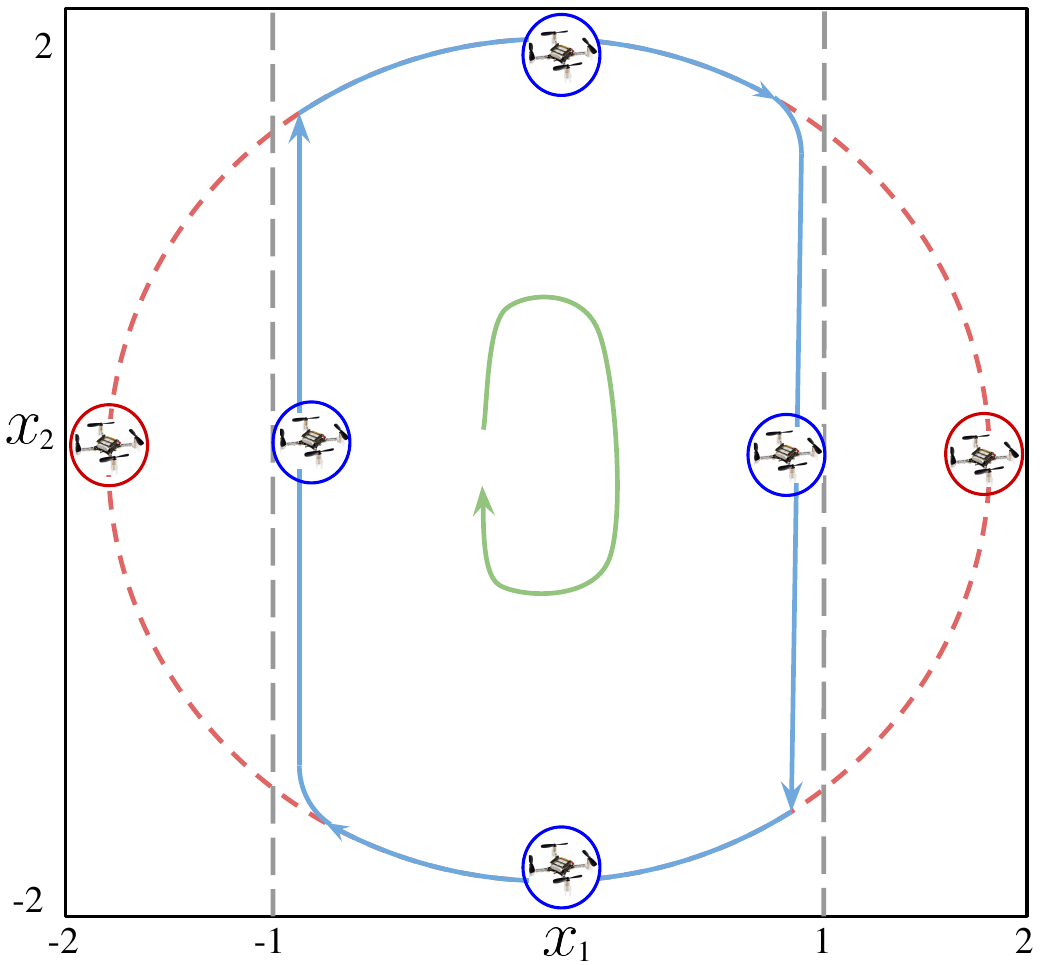}
    \caption{Implementation of the CP‑NCBF‑QP controller on a Crazyflie drone in Pybullet. The red line denotes the desired reference path, grey lines indicate the safety limits at $x_1=-1$ and $x_1=1$, and the blue line represents the safe trajectory generated by our controller.}
    \label{fig: uav-pybullet}
\end{figure}

Figure~\ref{fig: uav-pybullet} illustrates these results: the red curve corresponds to the reference trajectory, the grey vertical lines at $x_1 = -1$ and $x_1 = 1$ mark the boundaries, and the blue curve shows the trajectory actually executed by the safe controller on a Crazyflie drone in simulation.

\subsection{Quadruped Navigation in Dynamic Environment}\label{appendix: quadruped}

The control-affine dynamics of a reduced order Quadruped moving in a dynamic environment is given as:
\begin{equation}
    \left[\begin{array}{l}
    \dot{x}_1 \\
    \dot{x}_2 \\
    \dot{\phi} \\
    \dot{x}_{o1} \\
    \dot{x}_{o2} \\
    \dot{v}_{o1} \\
    \dot{v}_{o2} \\
    \dot{r}
    \end{array}\right]=\left[\begin{array}{c}
    0 \\
    0 \\
    0 \\
    v_{o1} \\
    v_{o2} \\
    k_1 \\
    k_2 \\
    k_r
    \end{array}\right]+\begin{bmatrix}
        cos\phi & 0 \\
        \sin\phi & 0 \\
        0 & 1 \\
        0 & 0 \\
        0 & 0 \\
        0 & 0 \\
        0 & 0 \\
        0 & 0 
    \end{bmatrix}
    \left[\begin{array}{c}
    u_1 \\
    u_2
    \end{array}\right],
\end{equation}
where $[x_1, x_2, \phi, x_{o1}, x_{o2}, v_{o1}, v_{o2}, r]^T\in\mathcal{X}\subseteq\mathbb{R}^8$ represents the state vector comprising the position $(x_1, x_2)$ and orientation $\phi$ of the quadruped, $(x_{o1}, x_{o2})$ represents the position of the obstacle, $(v_{o1}, v_{o2})$ represents the velocity of the obstacle and $r$ represents the radius of the obstacle. The inputs $u_1$ and $u_2$ correspond to the speed and the rate of change of orientation of the quadruped, respectively. 

The state space is defined as $\mathcal{X}= [-2,2]^2 \times [-\pi, \pi] \times [-2,2]^2\times [-1,1]^2\times [0.5,1]$. In this case we sample the safe and unsafe states using the concept of collision cone ~\citep{Chakravarthy2012, 709600}.

\begin{figure}
    \centering
    \includegraphics[width=1.0\linewidth]{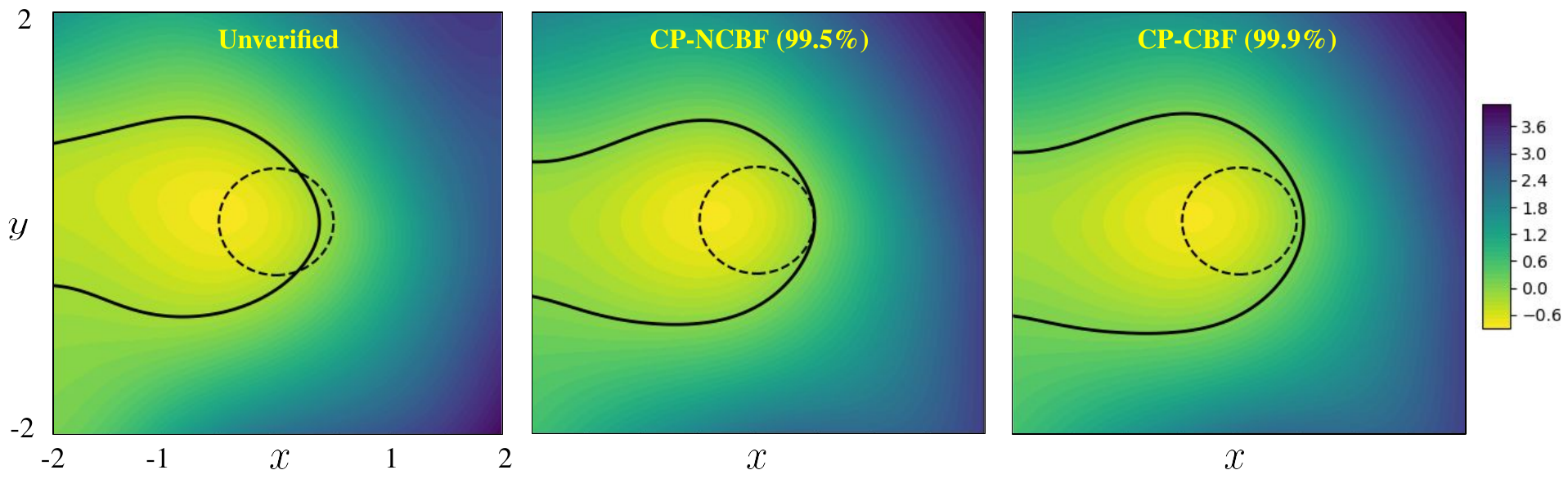}
    \caption{Visualization of learned NCBFs for the quadruped navigation at $\phi, x_{o1}, x_{o2}, v_{o1}, v_{o2} = 0$, $r=0.5$: (Left) Unverified NCBF (without robust loss), (Center) CP-NCBF (with 99.5\% safety guarantee), and (Right) CP-NCBF (with 99.9\% safety guarantee). Bold black lines denote the zero-level sets.}
    \label{fig: quad_nav}
\end{figure}

\subsubsection{Comparison of CP-NCBF with Baselines}

The Figure~\ref{fig: quad_nav} shows the visualization of the learned NCBFs for quadruped navigation in dynamic environments.

\subsubsection{Simulation Results}
We evaluated the CP‑NCBF‑QP controller (Eq.~\eqref{eq: CBF_QP}) at a $99.5\%$ safety level in a quadruped locomotion task implemented on the Pybullet physics engine \citep{coumans2019} via the Motion Imitation repository\footnote{\url{https://github.com/erwincoumans/motion_imitation}}. In this setup, the quadruped is commanded to follow a straight‐line trajectory while dynamically avoiding a moving obstacle.

\begin{figure}
\centering
\includegraphics[width=1.0\linewidth]{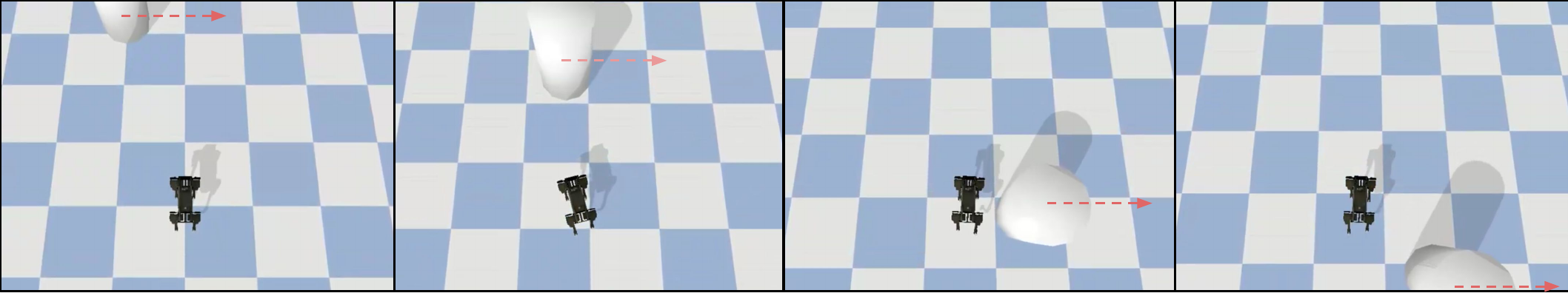}
\caption{Snapshots of quadruped navigation in Pybullet with a moving obstacle. The CP‑NCBF‑QP controller successfully prevents collisions while the agent progresses along its straight‐line path.}
\label{fig: quadruped-pybullet}
\end{figure}

Figure~\ref{fig: quadruped-pybullet} presents a sequence of snapshots capturing the quadruped as it safely negotiates the moving obstacle under the CP‑NCBF‑QP controller’s guidance.

\section{Implementation Details of the Algorithms}\label{appendix: implementation_details}

This section provides an in-depth overview of our algorithm and baseline implementations, including hyperparameter configurations and the cost/reward functions used in the baselines across all experiments.

\subsection{Experimental Hardware}
All experiments were conducted on a system equipped with an 11th Gen Intel Core i9-11900K @ 3.50GHz × 16 CPU, 128GB RAM, and an NVIDIA GeForce RTX 4090 GPU for training.

\subsection{Hyperparameters for the Proposed Algorithm}

The relative importance weights $\lambda_1, \lambda_2$ (in eq.\ref{eq: ncbf_loss}) are $1$ and $0.1$, respectively. 
The training settings for all the  experiments are detailed in the Table~\ref{tab:training_details}.

\begin{table}[h]
    \centering
    \begin{tabular}{lc}
        \hline
        \textbf{Hyperparameter} & \textbf{Value} \\
        \hline
        Network Architecture & Multi-Layer Perceptron (MLP) \\
        Activation Function & Softplus ($\log(1+\exp(x))$) \\
        Optimizer & Adam optimizer \\
        Learning Rate & $1\times 10^{-3}$ \\
        \hline
        \textbf{Autonomous Ground Vehicle Collision Avoidance} & \\
        \hline
        Number of Hidden Layers & 1 \\
        Hidden Layer Size & 64 neurons per layer \\
        Number of Training Points & 20K \\
        No. of Training Epochs & 1000 \\
        \hline
        \textbf{Aerial Vehicle Geo-Fencing} & \\
        \hline
        Number of Hidden Layers & 2 \\
        Hidden Layer Size & 128 neurons per layer \\
        Number of Training Points & 20K \\
        No. of Training Epochs & 1500 \\
        \hline
        \textbf{Quadruped Navigation in dynamic environment} & \\
        \hline
        Number of Hidden Layers & 2 \\
        Hidden Layer Size & 128 neurons per layer \\
        Number of Training Points & 50K \\
        No. of Training Epochs & 5000 \\
        \hline
    
    \end{tabular}
    \caption{Hyperparameters for the proposed algorithm}
    \label{tab:training_details}
\end{table}

\subsection{Hyperparameters for the Baseline } 
For the Lipschitz‑constrained NCBF method of \citet{tayal2024learning} (designed for autonomous ground vehicle collision avoidance), we adopted their publicly available implementation\footnote{\url{https://github.com/tayalmanan28/Stochastic-NCBF}} and set $\sigma=0$ to enforce the deterministic regime. Training required $N = 54\text{M}$ data points sampled densely across the state–action space; this sampling strategy, however, proved intractable for higher‑dimensional tasks such as aerial vehicle geofencing and dynamic quadruped navigation.

We also implemented the unverified neural CBF formulation of \citet{dawson2022safe}, using their open‑source codebase\footnote{\url{https://github.com/MIT-REALM/neural_clbf}}. In this case, we matched all hyperparameters to those specified for our approach (see Table~1 for details).

\subsection{Conformal Scores vs.\ Training Set Size}

We conducted an ablation study to investigate how the number of training samples affects the conformal safety guarantees of NCBFs in two settings: autonomous ground vehicle collision avoidance and aerial vehicle geo‑fencing.  

\begin{figure}
  \centering
  \begin{subfigure}[b]{0.77\linewidth}
    \centering
    \includegraphics[width=\linewidth]{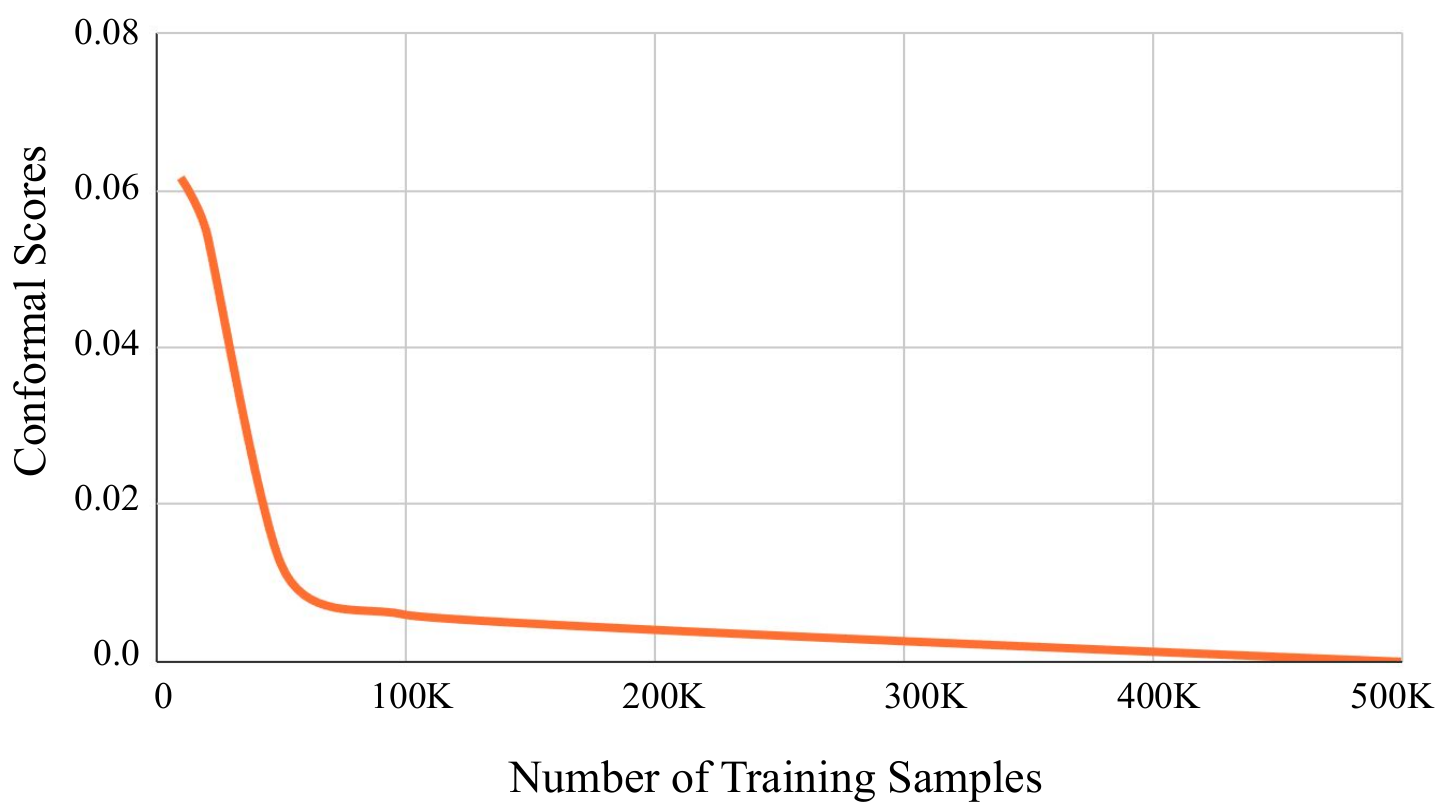}
    \caption{Ground vehicle: conformal score vs.\ number of training samples at $99.9\%$ safety.}
    \label{fig:verification-ground}
  \end{subfigure}
  \hfill
  \begin{subfigure}[b]{0.8\linewidth}
    \centering
    \includegraphics[width=\linewidth]{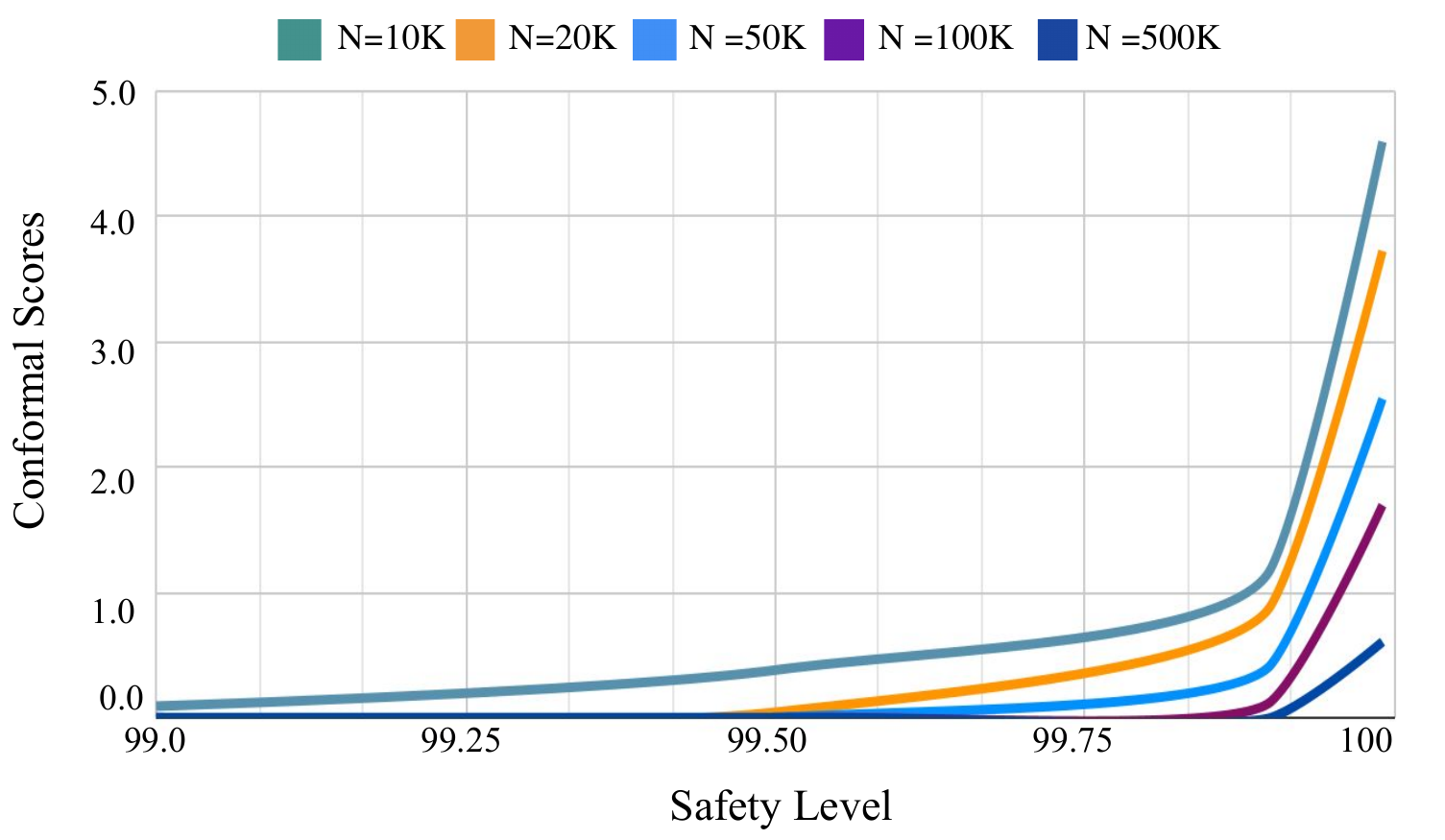}
    \caption{Aerial vehicle: conformal score vs.\ training samples for various safety levels.}
    \label{fig:verification-aerial}
  \end{subfigure}
  \caption{Dependence of conformal safety scores on the size of the training dataset for (a) ground‑vehicle collision avoidance and (b) aerial‑vehicle geo‑fencing tasks.}
  \label{fig:verification}
\end{figure}

Figure~\ref{fig:verification} summarizes our findings.  In subfigure~\ref{fig:verification-ground}, we plot the conformal scores of ground‑vehicle NCBFs trained on varying dataset sizes, all evaluated at the $99.9\%$ safety level.  Subfigure~\ref{fig:verification-aerial} shows the conformal scores for aerial‑vehicle NCBFs as a function of training sample size and different target safety levels. It can be inferred that the conformal scores decrease as the sample size increases and rise when stricter safety guarantees are imposed.

\end{document}